\documentclass[showpacs,superscriptaddress,nofootinbib,twocolumn]{revtex4}
\usepackage{graphicx}
\newcommand{\be}{\begin{equation}}
\newcommand{\ee}{\end{equation}}
\newcommand{\ba}{\begin{eqnarray}}
\newcommand{\ea}{\end{eqnarray}}
\newcommand{\la}{\langle}
\newcommand{\ra}{\rangle}

\newcommand{\di}{{\rm d}}
\newcommand{\sigmaPiN}{\sigma_{\pi N}}
\newcommand{\gsim}{{\renewcommand{\arraystretch}{0.35}\begin{array}{c}
   \mbox{\footnotesize$>$}\cr\mbox{\footnotesize$\sim$} \end{array}}}

\begin{document}
\title{\boldmath
	Extraction of the pion-nucleon sigma-term\\
	from the spectrum of exotic baryons}
\author{P.~Schweitzer}
\affiliation{Institut f\"ur Theoretische Physik II,
 	Ruhr-Universit\"at Bochum,  D-44780 Bochum, Germany}
\date{April 2004}
\begin{abstract}
  The pion nucleon sigma-term is extracted on the basis of the soliton 
  picture of the nucleon from the mass spectrum of usual and the recently 
  observed exotic baryons, assuming that they have positive parity. The value 
  found is consistent with that inferred by means of conventional methods from
  pion nucleon scattering data. The study can also be considered as a 
  phenomenological consistency check of the soliton picture of baryons.
\end{abstract} 

% 12.38.Lg = Other nonperturbative calculations
% 12.39.Fe = Chiral Lagrangians
% 12.39.Ki = Relativistic quark model
% 13.60.-r = Photon and charged-lepton interactions with hadrons	
% 14.20.-c   Baryons (including antiparticles) 
% 14.20.Dh = Protons and neutrons
%
%\keywords{Suggested keywords}
\maketitle

%====== SECTION 1: INTRODUCTION ====================================
\section{Introduction}

No experimental method is known to directly measure the pion-nucleon 
sigma-term $\sigmaPiN$ \cite{review-early,review-recent}. An indirect method
consists in exploring a low energy theorem \cite{low-energy-theorem} which 
relates the value of the scalar-isoscalar form factor $\sigma(t)$ at the 
point $t=2m_\pi^2$ to the isospin-even pion-nucleon scattering amplitude.
Earlier analyses by Koch \cite{Koch:pu} and Gasser et al.\ \cite{Gasser:1990ce}
gave for $\sigma(2m_\pi^2)$ a value about $60\,{\rm MeV}$, cf.\ Fig.~1.
From the difference $\sigma(2m_\pi^2)-\sigma(0)$ found to be $15\,{\rm MeV}$
by Gasser et al.\  \cite{Gasser:1990ce} one obtains for 
$\sigmaPiN\equiv\sigma(0)$ a value about $45\,{\rm MeV}$ 
which was generally accepted until the late 1990's.

Recent analyses \cite{Kaufmann:dd,Olsson:1999jt,Pavan:2001wz}, however, 
tend to yield higher values for $\sigma(2m_\pi^2)$ in the range 
$(80-90)\,{\rm MeV}$, cf.\ Fig.~1, due to the impact of more recent and 
accurate data \cite{Olsson:pi}.
This results in a value of $\sigmaPiN$ around $70\,{\rm MeV}$. 
Such a large value of $\sigmaPiN$ causes puzzles. 
According to a standard interpretation it implies a surprizingly large 
strangeness content of the nucleon (defined below in Eq.~(\ref{Def:y})), 
in contrast to what one would expect on the basis of the OZI-rule.

The precise knowledge of the value of $\sigmaPiN$ is, however, of practical 
importance for numerous phenomenological applications. E.g., the value of 
$\sigmaPiN$ enters the estimates of counting rates in searches of the Higgs 
boson \cite{Cheng:1988cz}, supersymmetric particles \cite{Bottino:1999ei} or 
dark matter \cite{Chattopadhyay:2001va,Prezeau:2003sv}. Therefore 
independent and direct methods to access $\sigmaPiN$ are welcome. 

In this note we would like to draw the attention to a method relying on the
soliton picture of the nucleon. The idea that baryons are different rotational 
excitations of the same object -- a classical soliton of the chiral field -- 
leads to numerous phenomenological relations among observables of different 
baryons, which are satisfied to a good accuracy and are model-independent, in 
the sense that they are due to symmetries of the soliton and do not depend 
on the dynamics of the respective model in which the soliton is realized.

For $\sigmaPiN$ no such model-independent relation could be found. 
All one can do -- sticking to known baryons -- is to relate $\sigmaPiN$ to 
mass splittings (among baryons in the SU(3)-flavour octet $J^P=\frac{1}{2}^+$)
and the a priori unknown strangeness content of the nucleon 
\cite{review-early,review-recent}.

In other words, if one considers $\frac{1}{2}^+$ octet and $\frac{3}{2}^+$ 
decuplet baryons and explores soliton symmetries the information content 
is not sufficient to pin down $\sigmaPiN$.  The situation changes by 
including baryons from the next multiplet suggested by the soliton picture 
-- the $\frac{1}{2}^+$ antidecuplet. 
After the prediction of its mass and width by Diakonov, Petrov and Polyakov 
\cite{Diakonov:1997mm} a candidate for the exotic ``pentaquark'' $\Theta^+$, 
the lightest member of the antidecuplet, was observed by several groups 
\cite{Nakano:2003qx,Barmin:2003vv,Kubarovsky:2003nn,Stepanyan:2003qr,Asratyan:2003cb,Kubarovsky:2003fi,Airapetian:2003ri,Aleev:2004sa}.
More recently also the finding of the second exotic baryon $\Xi_{3/2}^{++}$
was reported \cite{Alt:2003vb}.

In fact, in the soliton picture of the nucleon in linear order in the strange 
quark mass the pion-nucleon sigma-term is unambiguously fixed in terms of mass
splittings among baryons in the octet, decuplet and antidecuplet.
Assuming that the exotic baryons $\Theta^+$ and $\Xi_{3/2}$ are members of the
antidecuplet allows to extract $\sigmaPiN$ from the spectrum of usual and 
exotic baryons. The result compares well to the value of $\sigmaPiN$ deduced 
from the more recent analyses of pion-nucleon scattering data. 
The quality and accuracy of such an extraction are discussed.

The note is organized as follows. In Sec.~\ref{Sec:sigma} $\sigmaPiN$ is 
introduced. Sec.~\ref{Sec:Baryons-in-the-soliton-picture} contains a brief 
description of the soliton picture of baryons.
In Sec.~\ref{Sec:exotic-baryons+sigma} the relation between 
baryon mass splittings and $\sigmaPiN$ is discussed. 
Sec.~\ref{Sec:conclusions} contains the conclusions.

%
%  BEGIN FIGURE 1: HISTORICAL DEVELOPMENT
% 
\begin{figure}
	\includegraphics[width=7.8cm,height=6cm]{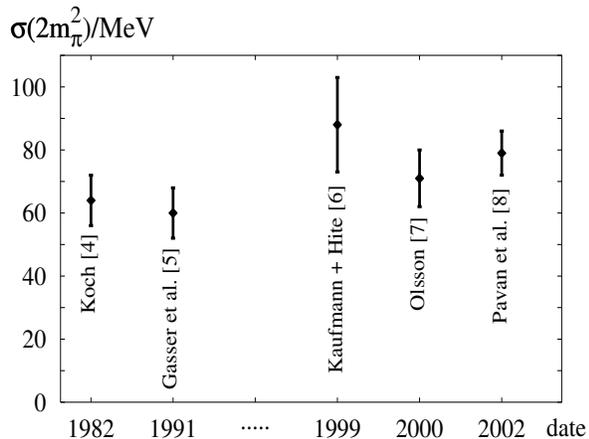}
	\caption{The ``historical development'' of the value for $\sigmaPiN$
	in the last two decades (the time-axis is not linear).}
\end{figure}
%
%  END FIGURE 1.
%

%====== SECTION 2: SIGMA-TERM ======================================
\section{The pion-nucleon sigma-term}
\label{Sec:sigma}

The nucleon sigma-term form factor $\sigma(t)$ and the pion-nucleon 
sigma-term $\sigmaPiN$ are defined as \cite{review-early,review-recent}
	\ba
 	&\sigma(t)\,\bar{u}({\bf p'})u({\bf p}) = 
	 m \,\la N'|\,(\bar{\psi}_u\psi_u+\bar{\psi}_d\psi_d)\,
	|N\ra \,,& \nonumber\\ 
	&& \nonumber\\
 	&\sigmaPiN = \sigma(t)|_{t=0}\;,\;\; 
	t = (p-p')^2,&
	\label{Eq:def-sigma(t)}\ea
where $m = \frac12(m_u+m_d)$ and the conventions are used
$\la N'|N\ra=2p^0\delta^{(3)}({\bf p}-{\bf p'})$ 
and $\bar{u}({\bf p})u({\bf p})=2M_N$. Strictly speaking in 
Eq.~(\ref{Eq:def-sigma(t)}) is neglected a ``doubly isospin violating term'' 
$\propto(m_u-m_d)(\bar{\psi}_u\psi_u-\bar{\psi}_d\psi_d)$. 

The form factor $\sigma(t)$ is a normalization scale invariant quantity,
which describes the elastic scattering off the nucleon due to the exchange 
of an isoscalar spin-zero particle. All that is known about it experimentally
is its value at the so-called Cheng-Dashen point $t=2m_\pi^2$.
A low energy theorem \cite{low-energy-theorem} relates $\sigma(2m_\pi^2)$ 
to the isospin-even pion nucleon scattering amplitude, which can be inferred from 
pion-pion and pion-nucleon scattering data by means of dispersion relations. 
Earlier analyses by Koch in 1982 \cite{Koch:pu} and Gasser et al.\ in 1991
\cite{Gasser:1990ce} gave, cf.\ Fig.~1, 
\be\label{sigma-2mpi^2-early}
	\sigma(2m_\pi^2) = \cases{
	(64\pm 8)\,{\rm MeV} & (1982) \cite{Koch:pu}\cr
	(60\pm 8)\,{\rm MeV} & (1991) \cite{Gasser:1990ce}.} \ee
Gasser et al.\ \cite{Gasser:1990ce} found from a dispersion relation analysis 
supplemented by chiral constraints 
\be\label{sigma-diff}
	\sigma(2m_\pi^2)-\sigma(0) = (15.2\pm 0.4)\;{\rm MeV} \;,
\ee
which gave for $\sigmaPiN$ a value about $45\,{\rm MeV}$. Modern analyses 
yield a larger value for the form factor at the Cheng-Dashen point
\be\label{sigma-2mpi^2-recent}
	\sigma(2m_\pi^2) = \cases{
	(88\pm15)\,{\rm MeV} & (1999) \cite{Kaufmann:dd}\cr
	(71\pm 9)\,{\rm MeV} & (2000) \cite{Olsson:1999jt}\cr
	(79\pm 7)\,{\rm MeV} & (2002) \cite{Pavan:2001wz}\cr
        (80\!-\!90)  \,{\rm MeV} & (2002) \cite{Olsson:pi},}\ee
which can be explained by the impact the more recent and accurate data 
\cite{Olsson:pi}. Thus recent analyses suggest
\be\label{sigmaPiN}
	\sigmaPiN \simeq (60-80)\,{\rm MeV} \;. \ee
The analyses of pion-nucleon scattering data are involved and it is difficult 
to control the systematic error both, of the extractions of $\sigma(2m_\pi^2)$
and its connection to $\sigmaPiN$ 
\cite{Koch:pu,Gasser:1990ce,Kaufmann:dd,Olsson:1999jt,Pavan:2001wz,Olsson:pi}.
However, there are no alternative methods to determine $\sigmaPiN$.

The sum rule $\sigmaPiN=m\int_0^1\!\di x\,(e^u+e^d+e^{\bar u}+e^{\bar d})(x)$ 
due to Jaffe and Ji \cite{Jaffe:1991kp}, which connects $\sigmaPiN$ to the 
chirally odd twist-3 nucleon distribution function $e^a(x)$, is unfortunately 
useless as an alternative method to learn about $\sigmaPiN$.
On top of practical difficulties to access chirally odd (and twist-3) 
distribution functions in deeply inelastic scattering experiments 
\cite{Avakian:2003pk}, there is also a theoretical obstacle prohibiting such a 
``measurement'' of $\sigmaPiN$.
The sum rule is saturated by a $\delta(x)$-type singularity. 
Such singularities can be (and were) ``observed'' in theoretical calculations 
\cite{Burkardt:1995ts,Schweitzer:2003uy} but in experiment they can manifest
themselves, in the best case, as a violation of the purely theoretical sum 
rule \cite{Efremov:2002qh}.

In lattice QCD -- the most direct approach to QCD -- the description of 
$\sigmaPiN$ is (at present) challenging. Direct lattice calculations of 
$\sigmaPiN$ meet the problem that the operator $\bar\psi\psi$ is not 
renormalization scale invariant \cite{Dong:1995ec}. An indirect method 
consists in exploring the Feynman-Hellmann theorem 
\cite{Feynman-Hellmann-theorem} 
\be\label{Eq:Feynman-Hellmann-II}
	\sigmaPiN = m\frac{\partial M_N}{\partial m}
	= m_\pi^2\frac{\partial M_N}{\partial m_\pi^2}
	\;,\ee
to deduce $\sigmaPiN$ from the pion mass dependence of the nucleon mass
measured on the lattice \cite{Bernard:2001av,AliKhan:2001tx,Zanotti:2001yb}.
In either case one faces the problem of extrapolating lattice data from
presently $m_\pi\gsim 500\,{\rm MeV}$ down to the physical value of the 
pion mass which is subject to systematic uncertainties which are difficult 
to estimate. Results of extrapolations of most recent and accurate lattice
data cover the range
\be\label{Eq:lattice}
	\sigmaPiN = (37^{+35}_{-13} - 73^{+15}_{-15})\,{\rm MeV}
	\ee
depending on the extrapolation ansatz \cite{Leinweber:2003dg}.
Chiral perturbation theory can in principle provide a rigorous guideline for 
the chiral extrapolation of lattice data -- provided one is able to control 
the convergence of the chiral expansion up to $m_\pi\gsim 500\,{\rm MeV}$
which seems feasible \cite{Bernard:2003rp,Procura:2003ig}.
Chiral perturbation theory does not, however, allow to compute $\sigmaPiN$ 
itself, which serves to absorb counter terms and has to be renormalized 
anew in each order of the chiral expansion.

The pion nucleon sigma-term was discussed in numerous models. See, e.g., 
\cite{Schweitzer:2003sb,Lyubovitskij:2000sf} for overviews of more recent 
works.

%====== SECTION 3: BARYONS AS SOLITONS ====================================
\section{Baryons in the soliton picture}
\label{Sec:Baryons-in-the-soliton-picture}

Since the early days of hadron physics symmetry principles have provided 
powerful guidelines for the qualitative classification of hadrons and 
the quantitative understanding of the hadron mass spectrum.

In this context it is worthwhile to recall the relations derived by
Gell-Mann and Okubo \cite{Gell-Mann,Okubo:1961jc} by considering SU(3) 
flavour symmetry and its breaking by quark mass terms up to linear order,
\ba\label{Gell-Mann-Okubo-I} &&	2M_{\rm N}+2M_{\Xi}=3M_\Lambda+M_\Sigma \;,\\
   \label{Gell-Mann-Okubo-II}&&	M_\Delta    -M_{\Sigma^\ast} = 
   M_{\Sigma^\ast}-M_{\Xi^\ast}=M_{\Xi^\ast}-M_\Omega \;,
\ea
which are full-filled to within few percent. 
Historically Eq.~(\ref{Gell-Mann-Okubo-II}) was used to predict the 
mass of the $\Omega^-$ baryon with impressive accuracy \cite{Eight-fold-way}.

The Gell-Mann--Okubo formulae relate baryon masses within a multiplet, 
namely the octet in Eq.~(\ref{Gell-Mann-Okubo-I}) and the decuplet in 
Eq.~(\ref{Gell-Mann-Okubo-II}), cf.\  Figs.~1a and 1b.
In order to relate masses from different multiplets one needs, however, 
more than the assumption of flavour symmetry. The limit of a large number 
of colours $N_c$ -- first discussed by 't Hooft \cite{'tHooft:1973jz} 
-- provides further symmetry arguments. 

Though in nature $N_c=3$ seems not to be large the multi-colour limit yields
numerous phenomenologically successful relations \cite{Jenkins:1998wy}.
In particular the large $N_c$ limit provides the basis for the picture of the 
nucleon as a classical soliton of the chiral pion field \cite{Witten:1979kh}. 
In the Skyrme model \cite{Skyrme:vq} or the chiral quark-soliton 
model \cite{Diakonov:1987ty} this picture is practically realized. 

In these models the nucleon is a soliton of the pion field 
$U_{{\rm SU}(2)}=\exp(i\tau^a\pi^a)$ which is of the so-called hedgehog shape
\be\label{Eq:hedgehog}
	\pi^a({\bf x}) = 
	\frac{\,{\rm x}^a\!}{|{\bf x}|}\,P\left(|{\bf x}|\right) \;,
\ee
such that flavour and space rotations become equivalent. Flavour SU(3) 
symmetry is considered by means of the following ``embedding'' ansatz 
\be\label{Eq:SU(3)-embedding}
	U_{{\rm SU}(3)} = 
	\left(\matrix{ U_{{\rm SU}(2)}\!\!\! & \matrix{0\cr 0} \cr
	0\;\;\;\;0 & 1}\right) \;.
\ee
In order to provide the classical soliton with spin, isospin and 
strangeness quantum numbers one has to consider the rotated field
\be\label{Eq:rotation}
	U_{{\rm SU}(3)}({\bf x},t)=R(t)\,U_{{\rm SU}(3)}({\bf x})\,R^\dag(t)
\ee
with $R(t)$ a time-dependent unitary SU(3) matrix. The quantization of the 
soliton rotation leads to the following rotational Hamiltonian and constraint
\be\label{Eq:H_rot}
H_{\rm rot} = \frac{1}{2I_A}\sum_{a=1}^3 J_a^2
	    + \frac{1}{2I_B}\sum_{a=4}^7 J_a^2 \;,\;\;\;
	J_8 = -\,\frac{N_c B}{2\sqrt{3}} \;.
\ee
In Eq.~(\ref{Eq:H_rot}) the $J_a$ ($a=1,\,2,\,\dots, 8$) are the generators
of the SU(3) group and $I_A$, $I_B$ are moments of inertia characterizing the 
rotation of the soliton. The eigenfunctions of $H_{\rm rot}$ -- the rotational baryon 
wave-functions with definite  spin, isospin and strangeness quantum numbers -- 
can be expressed in terms of Wigner finite-rotation matrices.

Of importance is the constraint of the generator $J_8$ in terms of the 
baryon number $B=1$. In the Skyrme model it is due to the Wess-Zumino term 
\cite{Witten:1979kh,Guadagnini:uv}. In the chiral quark soliton model it 
arises from a discrete bound state level in the spectrum of the single-quark
Hamiltonian in the background of the static soliton field \cite{Blotz:1992pw}.
The consequence of this constraint is that only SU(3) multiplets containing 
particles with hypercharge $Y=1$ are allowed. The lowest multiplets 
are the octet and decuplet of $J^P={\frac{1}{2}}^+$ and ${\frac{3}{2}}^+$ 
baryons respectively, cf.\  Figs.~1a and 1b.

In order to describe mass splittings within different multiplets
it is necessary to introduce explicit chiral symmetry breaking by 
quark mass terms $\propto{\rm tr}\,\hat{m}(U-1)$ in the Skyrme model 
or $\psi\hat{m}\psi$ in the chiral quark-soliton model, where $\hat{m}$ is 
the SU(3) quark mass matrix with $m_u=m_d = 0$ and $m_s > 0$ in the following.

%
%  BEGIN FIGURE 2: MULTIPLETS
% 
\begin{figure*}[t!]\label{Fig:predictions-AUT-HERMES}
	\vspace{-1cm}
	\includegraphics[width=13cm,height=7cm]{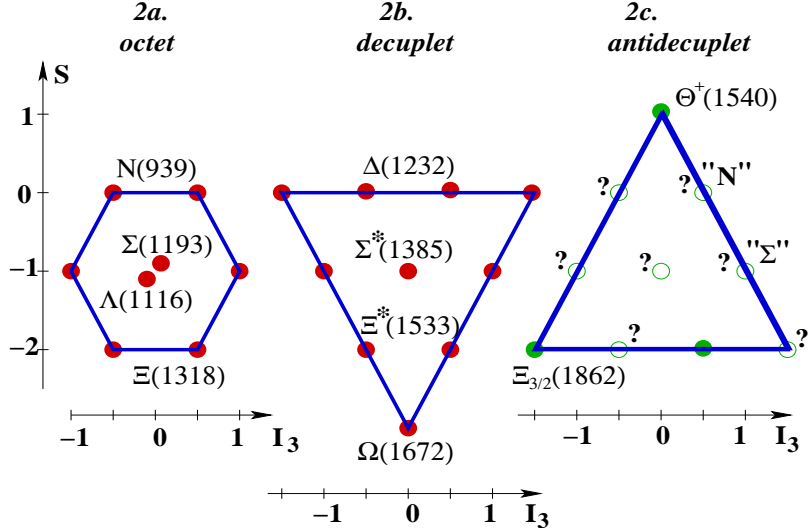}
	\caption{
	Baryon multiplets.
	{\bf a}. The $J^P=\frac12^+$ octet. 
	{\bf b}. The $J^P=\frac32^+$ decuplet.
	{\bf c}. The $J^P=\frac12^+$ antidecuplet predicted in the soliton 
	picture of the nucleon with the two recently observed exotic 
	candidates. The numbers in brackets denote the baryon masses 
	(averaged over isospin where necessary) in ${\rm MeV}$.}
\end{figure*}
%
%  END FIGURE 2.
%

The exploration of the spin-flavour symmetry of the rotating soliton and 
its explicit breaking by linear quark masses terms yields relations among 
observables of different baryons, which are well satisfied in nature and 
model-independent --  in the sense that they follow from symmetry 
considerations alone and do not depend on details of the dynamics, i.e.\  
on how and in which theory the self-consistent field $U$ is determined 
\cite{Adkins:1983ya,Adkins:cf,Guadagnini:uv}.

In particular one finds that (for $m_u = m_d = 0$) the eight different baryon 
masses in the octet and decuplet can be described in terms of 4 parameters: 
\begin{itemize}
\item{2 parameters fix the mass splittings within the multiplets,}
\item{1 parameter characterizes the mass splitting between octet and decuplet,}
\item{1 parameter fixes the absolute mass scale for one multiplet, 
      cf.\ \cite{Guadagnini:uv}.}
\end{itemize}
These parameters can, of course, be computed in a specific model. However, 
what is more interesting in our context is to find general model-independent 
(in the above sense) relations, which allow to test phenomenologically the 
underlying idea of the soliton symmetry.

By eliminating the 4 parameters one obtains 4 relations among 
the eight baryon masses, namely the 3 Gell-Mann--Okubo formulae,  
Eqs.~(\ref{Gell-Mann-Okubo-I},~\ref{Gell-Mann-Okubo-II}), and
in addition the Guadagnini relation \cite{Guadagnini:uv}
\be\label{Eq:Guadagnini}
	8(M_{\Xi^\ast}-M_{\Sigma^\ast}) = 11 M_\Lambda-8M_N-3M_\Sigma\,.
\ee 
Eq.~(\ref{Eq:Guadagnini}) relates mass splittings within different multiplets 
to each other. It is satisfied to an impressive accuracy of $1\%$.

%====== SECTION 4: EXOTIC BARYONS ==================================
\section{Exotic baryons and the pion-nucleon sigma-term}
\label{Sec:exotic-baryons+sigma} 

The soliton symmetry as described by means of the rotational Hamiltonian
(\ref{Eq:H_rot}) allows also higher multiplets.  The next multiplet, after 
the octet and decuplet, is the $J^P={\frac{1}{2}}^+$ antidecuplet, see Fig.~1c,
which contains new ``exotic'' baryons. 

From the point of view of the soliton picture there is nothing ``unusual'' 
about the baryons referred to as $\Theta^+$ and $\Xi_{3/2}$.
In a quark model, however, their quantum numbers can only be constructed
by including an additional $\bar qq$ pair. $\Theta^+$ has isospin 
zero and strangeness $S=1$ which requires a combination $uudd\bar s$.
The $\Xi_{3/2}^{--}$ member of the $\Xi_{3/2}$ isospin-quadruplet has 
$S=-2$ and $I_3=-\frac32$ which requires $\bar uddss$, etc.

The other members of the antidecuplet, denoted here as $''\Sigma''$ and 
$''N''$, have ``usual'' (in the quark model language) quantum numbers. 
Candidates for these baryons were discussed in Ref.~\cite{Diakonov:2003jj}. 
An unambiguous identifications of these states is difficult since mixings of 
the group theoretical states $|''N''\ra$ and $|''\Sigma''\ra$ with resonances 
of otherwise identical quantum numbers can occur.
For our purposes it is important to note that to linear order in $m_s$ such 
mixings do not effect the mass splittings within the octet and antidecuplet 
\cite{Diakonov:2003jj}.

In the description of the masses of the antidecuplet 
(always to linear order of quark masses) two additional parameters appear:
\begin{itemize}
\item{one characterizes the mass splittings,}
\item{the other fixes the absolute mass scale.}
\end{itemize}
The situation can be summarized as follows
\ba
&&	M_N             = M_8 - 7A - B,\nonumber\\  
&&	M_\Lambda       = M_8 - 4A, 	\nonumber\\
&&	M_\Sigma        = M_8 + 4A,	\nonumber\\
&&	M_\Xi           = M_8 + 3A + B,	\label{Eq:summary-octet}\\ 
&& 					\nonumber\\
&&	M_\Delta        = M_{10}-B     ,\nonumber\\   
&&      M_{\Sigma^\ast} = M_{10}, 	\nonumber\\
&&    	M_{\Xi^\ast}    = M_{10}+B,     \nonumber\\  
&&  	M_\Omega \;     = M_{10}+2B    ,\label{Eq:summary-decuplet}\\ 
&& 					\nonumber\\
&&	M_{\Theta^+}\;  = M_{\overline{10}} - 2B + 2C,\nonumber\\	
&&	M_{''N''}       = M_{\overline{10}} -  B +  C,\nonumber\\
&&	M_{''\Xi''}     = M_{\overline{10}},	      \nonumber\\
&&	M_{\Sigma_{3/2}}= M_{\overline{10}} +  B -  C,
\ea
where $M_8$,  $M_{10}$ and $M_{\overline{10}}$ characterize the average mass 
of the respective multiplet and $A$, $B$, $C$ the splittings within the 
multiplets.

The 12 baryon masses can thus be expressed by means of 6 parameters. 
Eliminating these parameters one obtains in addition to the 3 
Gell-Mann--Okubo and Guadagnini formulae,
Eqs.~(\ref{Gell-Mann-Okubo-I},~\ref{Gell-Mann-Okubo-II}), 
two further relations.  
The new relations express an equal-mass splitting rule in the antidecuplet,
\be\label{Eq:spliting-antidec}
	M_{\Sigma_{3/2}} - M_{''\Xi''} = 
	M_{''\Xi''} - M_{''N''} =
	M_{''N''} - M_{\Theta^+}\;, \ee
which is analogous to the relation (\ref{Gell-Mann-Okubo-II}) in the decuplet 
and was also observed in a description of pentaquarks in chiral perturbation 
theory \cite{Ko:2003xx}.
Thus, neither the mass splitting in the new antidecuplet nor its absolute 
scale $M_{\overline{10}}$ can be fixed in terms of known baryon masses.

As observed by Diakonov, Petrov and Polyakov (this was actually an important
ingredient in the prediction of Ref.~\cite{Diakonov:1997mm}) the pion-nucleon
sigma-term can be expressed in terms of the same parameters, namely
\be\label{sigma-ABC}
	\frac{m_s}{m} \, \sigmaPiN = 3(35 A + B + 4C) \;.
\ee
At first glance one could be worried by the appearance of $m=\frac12(m_u+m_d)$ 
in the denominator Eq.~(\ref{sigma-ABC}) since we work here in the chiral 
limit for light quarks $m_u=m_d=0$. However, one has to recall that 
$\sigmaPiN/m$ has a well-defined chiral limit -- also in soliton models, 
see e.g.\  \cite{Schweitzer:2003sb}. 

Eliminating the constants $A$, $B$, $C$ in  Eq.~(\ref{sigma-ABC}) one obtains
\ba\label{sigma-masses}
	\frac{m_s}{m} \, \sigmaPiN &= 
	\underbrace{3(4M_\Sigma-3M_\Lambda-M_N)}_{\mbox{\sl octet}} 
      + \underbrace{4( M_\Omega-M_\Delta)}_{\mbox{\sl decuplet}}&\nonumber\\
     & -\underbrace{4(M_{\Xi_{3/2}}-M_{\Theta^+})}_{\mbox{\sl antidecuplet}}&
	\;. \ea
Thus, the soliton picture connects $\sigmaPiN$ directly to the spectrum 
of baryons. In linear order of $m_s$ the relation is simple but the prize 
to pay is that antidecuplet baryons are involved.
In principle, if the antidecuplet would be established, the relation 
(\ref{sigma-masses}) would provide an attractive method to extract $\sigmaPiN$. 

Several comments are in order. In chiral perturbation theory {\sl ratios} of 
quark masses can be considered as convention and scale independent quantities 
\cite{Leutwyler:1996qg}. The framework of chiral perturbation theory 
eventually allows to express ratios of quark masses in terms of meson masses.

Eq.~(\ref{sigma-masses}) follows from evaluating linear $m_s$ effects in 
the soliton model. Therefore, for sake of consistency, it is preferable to use 
the value $m_s/m \equiv 2m_s/(m_u+m_d)=25.9$ resulting from the consideration 
of chiral symmetry breaking effects to linear order in quark masses 
\cite{Weinberg:hb,Leutwyler:1996qg}. Quadratic corrections yield
$m_s/m=24.4\pm 1.5$ \cite{Leutwyler:1996qg} -- which is a small numerical 
change in view of the accuracy to which we work here. 

Strictly speaking in the above discussion in the soliton model $m_s\neq0$ but 
for light quarks the chiral limit was considered, i.e.\  $m=(m_u+m_d)/2=0$.
Thus Eq.~(\ref{sigma-masses}) gives the correct relation between
$m_s\lim_{m\to 0}\sigmaPiN/m$ on the left-hand-side and baryon mass splittings 
on the right-hand-side (in the limit $m\to 0$). If one wished to 
include finite-$m$ effects one should consider also corrections 
due to $m_u\neq m_d$ and electromagnetic interactions on the same footing, 
which are of comparable magnitude.  In principle the effect of such 
corrections can be minimized by considering particular linear combinations 
of masses from isospin multiplets instead of their averages as we do. 
However, for our purposes such corrections can be disregarded. 
When deducing $\sigmaPiN$ from Eq.~(\ref{sigma-masses}) we shall 
assume that the $\sigmaPiN/m$ varies little in the chiral limit.

In the literature  it is currently being debated whether the description of 
exotic baryons in the framework of soliton models can fully be justified in 
the large $N_c$-limit
\cite{Cohen:2003nk,Diakonov:2003ei,Itzhaki:2003nr,Pobylitsa:2003ju,Cohen:2003mc}.
It was argued that -- from the large $N_c$-limit point of view -- a consistent 
description of multiplets containing exotics requires to go beyond the rotating
soliton: For exotic multiplets vibrational modes may play an equally important
role, in contrast to the usual octet and decuplet.
Eq.~(\ref{sigma-masses}) relates $\sigmaPiN$ to mass splittings {\sl within} 
multiplets. The rotating soliton description of mass splittings {\sl within} 
multiplets could still be consistent with large $N_c$, e.g., when vibrational 
soliton modes were flavour independent or negligibly small with respect to the 
rotational zero modes. Then Eq.~(\ref{sigma-masses}) would be consistent also 
from the large-$N_c$ point of view. This issue, of course, deserves further 
investigations.

Eq.~(\ref{sigma-masses}) can be rewritten by adding arbitrary multiples of 
the following ``zeros'' 
\ba
	11 M_\Lambda+8(M_{\Sigma^\ast}-M_N-M_{\Xi^\ast})-3M_\Sigma = 0\,,
	&&\label{Eq:Guadagnini-zero}\\
	2M_{\Xi^\ast}- M_\Omega - M_{\Sigma^\ast} = 0\,,
	&&\label{Eq:Gell-Mann-Okubo-zero-2a}\\
	M_{\Xi^\ast}  + M_{\Sigma^\ast} -M_\Omega  -  M_\Delta = 0\,,
	&&\label{Eq:Gell-Mann-Okubo-zero-2b}\\
	3M_\Lambda-2M_{\rm N}-2M_{\Xi}+M_\Sigma = 0\,,
	&&\label{Eq:Gell-Mann-Okubo-zero}
\ea
which result from
Eqs.~(\ref{Gell-Mann-Okubo-I},~\ref{Gell-Mann-Okubo-II},~\ref{Eq:Guadagnini}).
Formally this would not change Eq.~(\ref{sigma-masses}). In practice, however, 
the ``zeros'' are only approximate.

Eq.~(\ref{Eq:Guadagnini-zero}) is the most exact ``zero'', the right-hand-side
(RHS) of Eq.~(\ref{Eq:Guadagnini-zero}) is $1\,{\rm MeV}$ if we insert baryon 
masses (averaged over isospin). Thus, if we added $25\times$ this 
``zero'' we would change the value of $\sigmaPiN$ by $1\,{\rm MeV}$ only. 
A common sense agreement could be to use Eq.~(\ref{Eq:Guadagnini-zero}) such 
that octet and decuplet (and uncertainties in their description) contribute 
to $\sigmaPiN$ with comparable weight.
Eq.~(\ref{sigma-masses}) represents a possible choice -- under the 
asthetical constraint to avoid awkward fractional coefficients.

Eqs.~(\ref{Eq:Gell-Mann-Okubo-zero-2a},~\ref{Eq:Gell-Mann-Okubo-zero-2b})
are less precise ``zeros''. The RHS of (\ref{Eq:Gell-Mann-Okubo-zero-2a})
is $9\,{\rm MeV}$ and the RHS of (\ref{Eq:Gell-Mann-Okubo-zero-2b}) yields
$14\,{\rm MeV}$. The uncertainty these relations introduce in 
Eq.~(\ref{sigma-masses}) can be estimated by using instead of 
$4(M_\Omega-M_\Delta)$, e.g., $12(M_{\Sigma^\ast}-M_\Delta)$ or 
$12(M_\Omega-M_{\Xi^\ast})$.
In this way one obtains $(1750\pm 90)\,{\rm MeV}$ for the contribution of the
decuplet in Eq.~(\ref{sigma-masses}). 
The RHS of Eq.~(\ref{Eq:Gell-Mann-Okubo-zero}) yields $27\,{\rm MeV}$.
We estimate the total contribution of the octet to Eq.~(\ref{sigma-masses}) 
as $(1455 \pm 150)\,{\rm MeV}$.

Turning to the antidecuplet let us first point out that by choosing exotic 
antidecuplet members in Eq.~(\ref{sigma-masses}) one avoids a principle 
complication, namely how to identify the non-exotic members in the new 
multiplet in view of possible complicated mixing patterns 
\cite{Diakonov:2003jj,Arndt:2003ga}.
(Recall that to linear order in $m_s$ mixing does not effect mass 
differences {\sl within} a multiplet.)
Taking the candidates for $\Theta^+$ and $\Xi_{3/2}$ for granted we obtain
for the contribution of the antidecuplet in Eq.~(\ref{sigma-masses}) the value
$(1288 \pm 150)\,{\rm MeV}$ presuming that the mass splitting formula in 
the antidecuplet works no better and no worse than in other multiplets.

Thus, we obtain for the pion-nucleon sigma-term 
\be\label{sigmaPiN-final}
	\sigmaPiN = (74 \pm 12)\,{\rm MeV} \:.
\ee
Alternatively, we can perform a best fit for the parameters $A$, $B$, $C$ 
in Eqs.~(\ref{Eq:summary-octet},~\ref{Eq:summary-decuplet}) which gives 
respectively 
$(9\pm  2)\,{\rm MeV}$, $(145\pm 12)\,{\rm MeV}$, $(37\pm 10)\,{\rm MeV}$. 
From (\ref{sigma-ABC}) one then obtains $\sigmaPiN = (71\pm 14)\,{\rm MeV}$, 
in agreement with (\ref{sigmaPiN-final}).
  (For completeness, the average masses of the multiplets are
   $M_8    	   = 1151 \,{\rm MeV}$,
   $M_{10} 	   = 1382 \,{\rm MeV}$,
   $M_{\overline{10}} = 1754 \,{\rm MeV}$.)

The result (\ref{sigmaPiN-final}) is in reasonable agreement with the 
value of $\sigmaPiN$ obtained from the recent dispersion relation analyses of 
pion nucleon scattering data, Eq.~(\ref{sigmaPiN}). It is also compatible with 
lattice results, Eq.~(\ref{Eq:lattice}).

Several comments are in order. 
Firstly, we took the candidates for $\Theta^+$ and $\Xi_{3/2}$ for granted.
However, in particular the $\Xi_{3/2}$ state has not yet been confirmed. 
Instead it was argued that the results of the NA49-experiment are in 
conflict with earlier experiments \cite{Fischer:2004qb}. 
Secondly, we assumed that the soliton picture describes the antidecuplet
to within the same accuracy as the octet and the decuplet, which can be
checked only after all (also non-exotic) members of the antidecuplet will
unambiguously be identified. 
Thirdly, the error in Eq.~(\ref{sigmaPiN-final}) reflects the accuracy 
to which the soliton picture describes the right-hand-side of 
Eq.~(\ref{sigma-masses}), which does not necessarily comprise the 
entire uncertainty to which the soliton relation (\ref{sigma-masses}) 
{\sl itself} is satisfied. The accuracy of (\ref{sigma-masses}) could be 
checked if we knew $\sigmaPiN$ (and all antidecuplet masses) precisely. 

Thus the error in (\ref{sigmaPiN-final}) could be underestimated. However, 
this error does not appear unrealistic in view of the experience with other 
soliton relations -- which connect, e.g., baryon-meson coupling constants 
\cite{Witten:1979kh}, magnetic moments \cite{Kim:1997ip} or hyperon decay 
constants \cite{Kim:1999uf}, and which typically hold to within an accuracy 
of $(10-20)\%$. (The Guadagnini formula (\ref{Eq:Guadagnini}) is another 
example.)

Finally, let us comment on the strangeness content of the nucleon which is 
defined as
\be\label{Def:y}
	y=\frac{2\la N|\bar\psi_s\psi_s|N\ra}
	       { \la N|\bar\psi_u\psi_u+\bar\psi_d\psi_d|N\ra}\;.\ee
The value of $y$ can be inferred from mass splittings of the octet baryons. 
To linear order quark masses one obtains 
\cite{review-early,review-recent}
\be\label{Eq:y-mass-splittings}
	y = 1 - \frac{m}{m_s-m}\;\frac{M_{\Xi}+M_{\Sigma}-2M_N}{\sigmaPiN}
	\;.\ee
By means of Eq.~(\ref{sigma-masses}) one can express $y$ entirely in terms 
of baryon mass splittings -- which yields $y\approx 0.6$. Inclusion of higher 
order quark mass terms tends to decrease the value of $y$ \cite{Gasser:1982ap}
-- which, however, still remains surprizingly large from the point of view of 
the OZI rule. The latter would imply the matrix element 
$\la N|\bar\psi_s\psi_s|N\ra$ to be small. 

The term ``strangeness content" is, however, somehow misleading. The 
scalar operator $\bar\psi_s\psi_s$ does not ``count'' strange quarks unlike 
the (zero component of the) vector operator $\bar\psi_s\gamma^\mu\psi_s$ does.
Thus strictly speaking there is no {\sl a priori} reason for the matrix 
element $\la N|\bar\psi_s\psi_s|N\ra$ to be small (apart from the OZI rule).
In spite of a large strangeness content $y$ the total contribution of 
strange quarks to the nucleon mass is reasonably small \cite{Ji:1994av}.

\section{Conclusions}
\label{Sec:conclusions}

In the soliton picture of baryons in the linear treatment of strange quark 
mass terms the pion-nucleon sigma-term is simply related to the mass 
splittings in the octet, decuplet and antidecuplet \cite{Diakonov:1997mm}.
Presuming that the $\Theta^+$ and $\Xi^{--}_{3/2}$ exotic baryons 
\cite{Nakano:2003qx,Barmin:2003vv,Kubarovsky:2003nn,Stepanyan:2003qr,Asratyan:2003cb,Kubarovsky:2003fi,Alt:2003vb,Airapetian:2003ri,Aleev:2004sa}
are members of the antidecuplet the pion-nucleon sigma-term was extracted 
from the mass splittings of usual and exotic baryons and found to be 
$\sigmaPiN=74\,{\rm MeV}$ with an accuracy of about $(15-20)\%$.
This result is in good agreement with recent analyses of pion-nucleon and 
pion-pion scattering data which yield for the scalar-isoscalar form factor 
at the Cheng-Dashen point $\sigma(2m_\pi^2)=(80-90)\,{\rm MeV}$
\cite{Kaufmann:dd,Olsson:1999jt,Pavan:2001wz,Olsson:pi}.

However, the present experimental basis for this analysis cannot be considered 
as solid. The $\Xi_{3/2}$ candidate has not yet been confirmed by independent 
groups, cf.\  Ref.~\cite{Fischer:2004qb} for a critical discussion.
The widths are not measured directly \cite{Theta+width}, and in particular 
spin and parity of the exotic baryons are not established \cite{Theta+parity}.
So it is not yet clear whether the exotic states fit into the soliton picture 
of the nucleon \cite{Diakonov:1997mm,Praszalowicz:1987} or into other 
approaches \cite{Jaffe:2003sg,Sasaki:2003gi,Shuryak:2003zi,Close:2003tv}.

If confirmed the soliton picture would provide an appealing method to access
$\sigmaPiN$ directly -- with an uncertainty comparable to the accuracy to 
which Gell-Man--Okubo, and Guadagnini mass relations are satisfied.
In future, with more information available on the antidecuplet, the 
uncertainty could be estimated more conservatively than it was possible here.
This method could provide valuable information on $\sigmaPiN$ supplementary 
to $\sigma(2m_\pi^2)$-extractions or direct lattice calculations.
At the present stage the exercize presented here can be considered as a 
consistency check of the soliton picture -- as it was presented along the 
lines of Ref.~\cite{Diakonov:2003jj}.

Further interesting issues are the inclusion of finite light quark current 
masses, isospin breaking effects or higher order strange quark mass corrections
by extending the methods elaborated in Ref.~\cite{Blotz:1992pw}.

\begin{acknowledgments}
The author thanks K.~Goeke for discussions and a careful reading of the 
manuscript, and is grateful to D.~I.~Diakonov for inspiring remarks.
This work is partially supported by Verbundforschung of BMBF.
\end{acknowledgments}

\end{document}